\documentclass{ws-jai}
\usepackage[flushleft]{threeparttable}
\usepackage{newtxtext,newtxmath}
\usepackage{indentfirst}
\usepackage{graphicx}
\usepackage{subfig}
\usepackage{subcaption}
\usepackage{multirow}
\usepackage{color}
\usepackage{booktabs}
\usepackage{footmisc}

\begin{document}

\catchline{}{}{}{}{} % Publisher's Area please ignore

\markboth{Wei Liu}{A new open-source frontend for CASPER toolflow}
\title{An Open-Source High-Level Graphical Signal Processing Language with Simulation and HDL Generation}
%\title{An Open-Source High-Level Graphical Signal Processing Language for FPGAs}
%\title{A new open-source frontend for CASPER toolflow}

\author{Wei Liu$^{1}$, Jonathon Kocz$^{1}$ and Dan Werthimer$^{1}$}

\address{
    $^{1}$Department of Astronomy, University of California, Berkeley, Berkeley, CA 94720, USA}

\maketitle
\corres{$^{1}$Wei Liu.}

\begin{history}
    \received{(to be inserted by publisher)};
    \revised{(to be inserted by publisher)};
    \accepted{(to be inserted by publisher)};
\end{history}a

\begin{abstract}
    The CASPER (Collaboration for Astronomy Signal Processing and Electronic Research) toolflow is a widely used framework for designing and implementing digital signal processing systems, particularly in the field of radio astronomy. It provides a set of tools and libraries that enable researchers to create custom hardware and software solutions for processing astronomical data. The CASPER toolflow has been instrumental in the development of Field-Programmable Gate Array (FPGA) based digital instruments for various radio telescopes, enabling for real-time data processing and analysis. However, the current frontend tool that CASPER uses for high-level FPGA design is based on Model Composer integrated into MATLAB/Simulink, which is a proprietary software. In this paper, we introduce Scilab as a new frontend tool for the CASPER toolflow. Scilab is an open-source software platform for numerical computation and data visualization, which offers a similar environment to MATLAB/Simulink for designing CASPER blocks, generating FPGA Intellectual Property (IP) cores, and simulating Digital Signal Processing (DSP) systems. We present our implementation of Scilab in the CASPER toolflow and demonstrate its capabilities by developing an FPGA based spectrometer on a RFSoC4x2, a commonly used CASPER platform well suited to radio astronomy applications. We have also developed Scilab support for other CASPER compatible platforms. Our results show that Scilab can successfully be used as an alternate frontend for CASPER-based designs.
\end{abstract}

\keywords{Techniques; Open Source; Instrumentation.}

\section{Introduction}

\indent Developed by the Collaboration for Astronomy Signal Processing and Electronics Research (CASPER\footnote{https://casper-astro.github.io}), the CASPER toolflow is an open-source framework designed to simplify and accelerate the development of FPGA-based Digital Signal Processing (DSP) systems \cite{hickish2016decade,parsons2009digital, parsons2008scalable, parsonsnew}. It provides a modular environment with reusable DSP libraries based on Simulink blocks that can be compiled for a range of supported hardware platforms. These hardware platforms (e.g., RFSoC4x2\footnote{https://casper-toolflow.readthedocs.io/projects/tutorials/en/latest/tutorials/rfsoc/platforms/rfsoc4x2.html}, RFSoC2x2\footnote{https://casper-toolflow.readthedocs.io/projects/tutorials/en/latest/tutorials/rfsoc/platforms/rfsoc2x2.html}, SNAP\footnote{https://casper-toolflow.readthedocs.io/projects/tutorials/en/latest/tutorials/snap/tut\_intro.html}, ROACH2\footnote{https://casper-toolflow.readthedocs.io/projects/tutorials/en/latest/tutorials/roach/tut\_intro.html} and SKARAB\footnote{https://casper-toolflow.readthedocs.io/en/latest/}) have pre-written interfaces (known as "Yellow blocks" in CASPER) that, combined with the DSP libraries, allow the user to develop instruments without needing a detailed understanding of FPGAs. This also abstracts away the need to learn the underlying MATLAB, Simulink, and AMD (Advanced Micro Devices, Inc.) Model Composer software that these libraries are based on. MATLAB is a numerical computing environment, Simulink is its graphical platform for modeling and simulating dynamic systems, and AMD Model Composer is an extension that enables model-based design and hardware acceleration on AMD Xilinx FPGAs and adaptive SoCs. This approach is particularly valuable to the radio astronomy community, where large-scale telescope arrays generate massive volumes of data requiring real-time processing. CASPER’s design ecosystem significantly reduces development time and lowers the barrier to entry for building complex instrumentation. \\

\indent The CASPER toolflow has been used to develop radio astronomy instrumentation that has been deployed at observatories worldwide. These instruments have yielded numerous scientific results. This includes such facilities and experiments as: the Event Horizon Telescope \cite{vertatschitsch2015r2dbe}; the Submillimeter Array \cite{johnson2015resolved}; the Hydrogen Epoch of Reionization Array (HERA) \cite{deboer2017hydrogen}; the Breakthrough Listen project \cite{macmahon2018breakthrough}; the Deep Synoptic Array 10 dish prototype and succeeding 110 dish array (DSA-10, DSA-110) \cite{kocz2019dsa, ravi2022detection}; the MeerKAT \cite{van2022meerkat} and the Five-hundred-meter Aperture Spherical Radio Telescope(FAST) \cite{zhang2020first}. \\

\indent While the current approach has been successful, compatibility issues and version constraints pose significant challenges for the community in adopting and collaboratively using the toolflow: Each version of Vivado is only compatible with specific versions of Model Composer, Matlab and operating systems. Additionally, the model libraries are not always backwards compatible. Therefore, we seek to improve these in this paper: 1) As CASPER depends on AMD Model Composer in MATLAB and Vivado, users must install both software packages on a specific operating system. Version compatibility is becoming a problem in the CASPER community as newer versions of MATLAB, Vivado, and the operating system are continuously released, which can delay the upgrade cycle of CASPER libraries. 2) CASPER has benefited from the use of MATLAB and Vivado. However, they remain closed-source, which, in addition to making it harder and harder to solve the version compatibility problem, is in opposition to the CASPER ethos of making all code open source for reuse by the community. \\

\indent In this paper, we introduce the use of an open-source frontend tool, Scilab, for CASPER. It provides a similar Graphical User Interface (GUI) and features to the user, making it easier for CASPER users to transition to the new environment. As Scilab is completely removed from Matlab and Model Composer, makes use of HDL libraries, and is not directly linked to Vivado, users will not be tied to the same version restrictions as with the current, Model Composer based toolflow. However, Scilab does not provide support for the Model Composer IP, nor the same integrated simulation capabilities. As such, all CASPER DSP blocks for use with the Scilab frontend have been recreated in Hardware Description Language (HDL). All of the HDL modules have been tested and compared with the Simulink modules, to ensure the functionality remains the same. Also, scripts to generate and run simulations of designs in Vivado have been created. This has the added benefit of easing the development and integration of other frontend interfaces (e.g., GNU Radio\cite{blossom2004gnu}) into the CASPER toolflow.\\

\indent In Section~\ref{casper-toolflow-workflow}, the detailed workflow of the current CASPER toolflow is introduced, including what blocks are provided in the frontend, and how the design is compiled; The new frontend, Scilab, is introduced in Section~\ref{casper_toolflow_in_scilab}, including how to create new Yellow/Green blocks in Scilab and how the new toolflow works; Section~\ref{other-frontend} gives a simple example of porting the CASPER toolflow in GNU Radio, which shows the use of other frontends; In Section~\ref{spectrometer-demo}, the improved toolflow is demonstrated by a spectrometer test design created using the new frontend and a comparison with the same design using the Simulink frontend. A summary of the current status and future directions of the work is given in Section~\ref{summary}. \\

\section{CASPER toolflow in MATLAB}
\label{casper-toolflow-workflow}

% \indent An FPGA design consists of two kinds of blocks: peripherals blocks and DSP blocks. The peripherals blocks are used to interface with the outside world, like using Analog-to-Digital Convertors(ADCs) to sample signals, using Digital-to-Analog Convertors(DACs) to output signals, using Ethernet module to transmit data, and so on. The DSP blocks are used to process the data from the peripherals or deliver the processed data to the peripherals in real time. A general FPGA design is shown in figure\ref{general_fpga_design}. \\
% \begin{figure}[htbp]
%     \centering
%     \includegraphics[height=2.3cm]{figures/general_fpga_design.png}
%     \caption{
%         A general FPGA design consists of input yellow blocks(like ADC blocks), green blocks(like Polyphase Filter Banks blocks, accumulation blocks.) and output yellow blocks(like 100GbE block).
%         \label{general_fpga_design}
%     }
% \end{figure}

\indent The CASPER toolflow has been successfully applied to numerous developments in radio astronomy instruments.\footnote{https://ui.adsabs.harvard.edu/public-libraries/7mm5XbqlRiW5FN-Si0l7oA}. It comprises two components, broken into the frontend and the backend. The frontend offers a graphical user interface together with a collection of peripheral (interface) and DSP blocks to facilitate user-friendly design creation. The backend compiles the design and generates a bitstream file for uploading to the hardware platform. The CASPER toolflow also offers simulation capabilities in the frontend for the designs, enabling users to verify their designs efficiently. All of the frontend functionality needs to be replicated in order for Scilab to be a powerful alternative. The specific requirements are detailed in subsections~\ref{casper-frontend} and \ref{casper-backend}.

\subsection{Frontend}
\label{casper-frontend}
\begin{figure}[htbp]
    \centering
    \includegraphics[height=1.5cm]{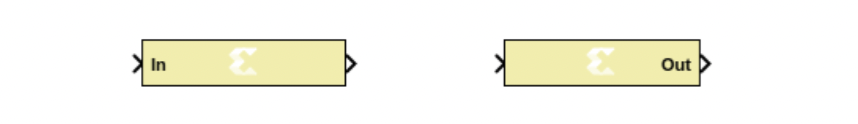}
    \caption{
        AMD Gateway In and Gateway Out blocks in Simulink. These could be multi-bit buses or single-bit lines, which depend on the design.
        \label{xilinx_gateway_blocks}
    }
\end{figure}

\begin{figure}[htbp]
    \centering
    \includegraphics[height=6cm]{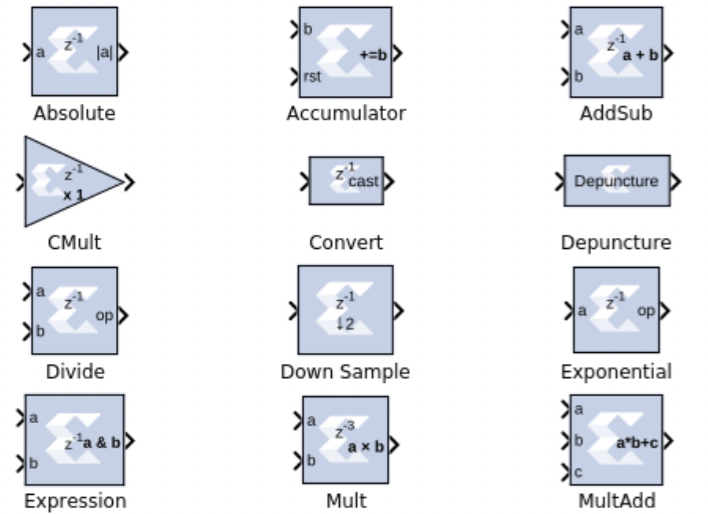}
    \caption{
        AMD Xilinx HDL Blockset in Simulink. (More AMD Xilinx basic blocks are found in the Simulink AMD library.)
        \label{xilinx_basic_blocks}
    }
\end{figure}

\begin{figure}[htbp]
    \centering
    \includegraphics[height=10cm]{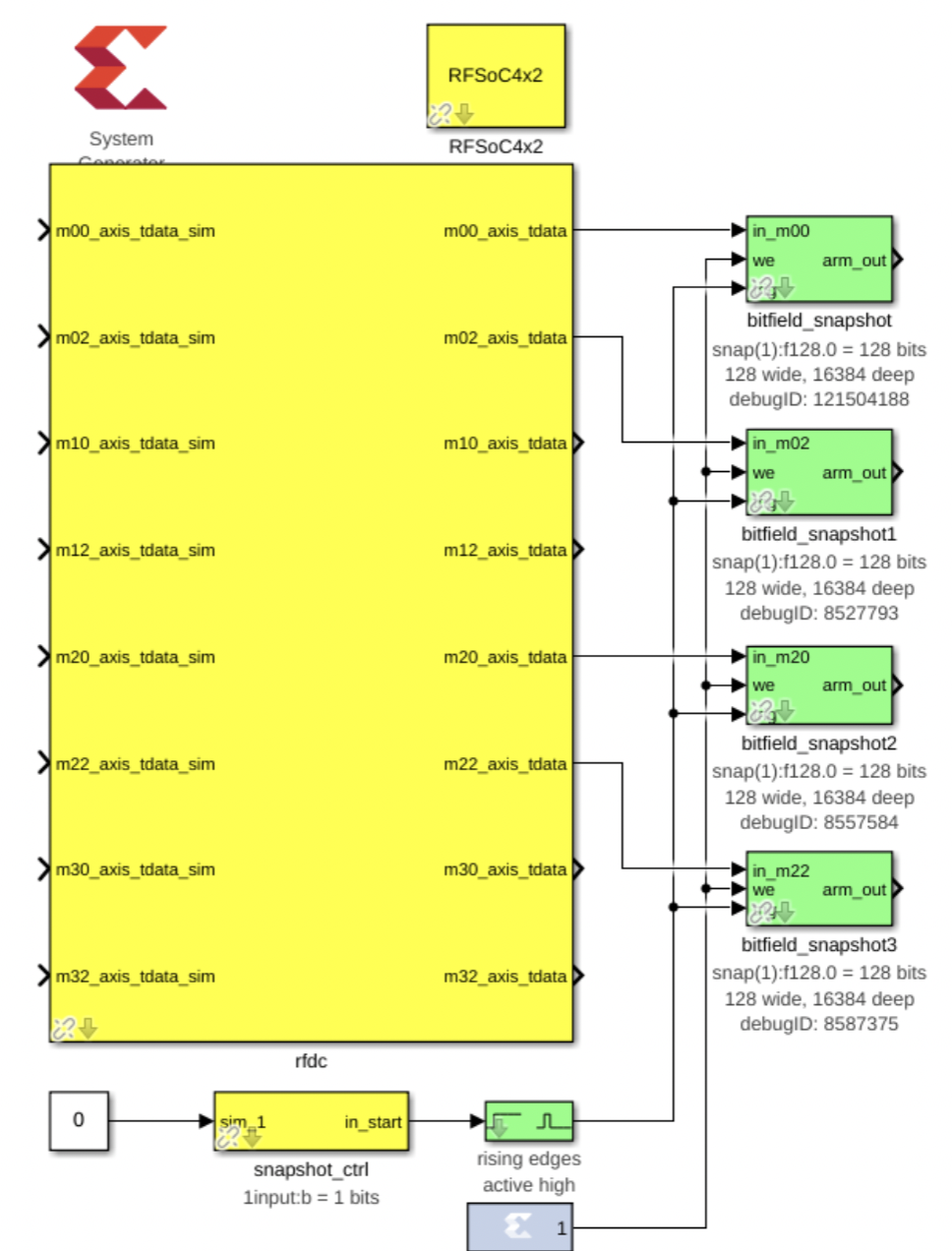}
    \caption{
        A CASPER example Simulink design targeting an RFSoC4x2 FPGA board. In this design, four snapshot blocks capture data from an RFDC (ADC) block. A software register and an edge detect block are used to enable/disable the data capture.
        \label{casper_example}
    }
\end{figure}

\indent The existing frontend tool relies on Simulink within MATLAB. Simulink is a graphical modeling platform created by MathWorks extensively utilized for system-level simulation, design, and automated code generation of dynamic and embedded systems. In Simulink, the CASPER toolflow offers Yellow blocks and Green blocks to users. Yellow blocks facilitate interaction with peripherals, such as ADCs, DACs, I/O, and memory. Various FPGA platforms are supported via distinct platform referred to as Yellow blocks, enabling the same design to be interoperable with multiple hardware configurations. The CASPER toolflow is currently compatible with 21 distinct FPGA systems. Green blocks are customized DSP components developed by CASPER, used within AMD Model Composer in MATLAB/Simulink for high-speed signal processing in radio  astronomy. The Green blocks comprise elements such as the Fast Fourier Transform (FFT), Polyphase Filter Bank (PFB), accumulators, and some additional, general purpose DSP blocks (e.g., complex multiplier). Most of the Yellow blocks are based on AMD Xilinx Gateway In/Out blocks, which are shown in Figure~\ref{xilinx_gateway_blocks}, while most of the Green blocks consist of a collection of ADM Xilinx HDL blocks\footnote{https://docs.amd.com/r/2021.1-English/ug1483-model-composer-sys-gen-user-guide/HDL-Blockset}, such as addition and multiplication. Some of these basic blocks are shown in Figure~\ref{xilinx_basic_blocks}. An example based on these CASPER blocks is shown in Figure~\ref{casper_example}. \\

% \indent CASPER blocks are fundamentally derived from Xilinx gateway and basic blocks, resulting in CASPER designs that predominantly incorporate these Xilinx blocks within Simulink. This is a model-based design tool incorporated into MATLAB/Simulink that facilitates the creation of FPGA-based digital signal processing (DSP) systems using a graphical interface. It offers a repository of Xilinx-specific blocks that enables designers to simulate, validate, and automatically create HDL code from Simulink models, thereby optimizing the deployment of intricate DSP algorithms on Xilinx FPGAs. In the CASPER design flow, XSG facilitates the amalgamation of low-level hardware components with advanced CASPER DSP modules, producing a DSP IP core for the designated FPGA platform. Model Composer block is show in figure\ref{system_generator_block}. \\

% \begin{figure}[htbp]
%     \centering
%     \includegraphics[height=9cm]{figures/system_generator_block.png}
%     \caption{
%         AMD Model Composer block in Simulink.
%         \label{system_generator_block}
%     }
% \end{figure}

\begin{figure}[htbp]
    \centering
    \includegraphics[height=5cm]{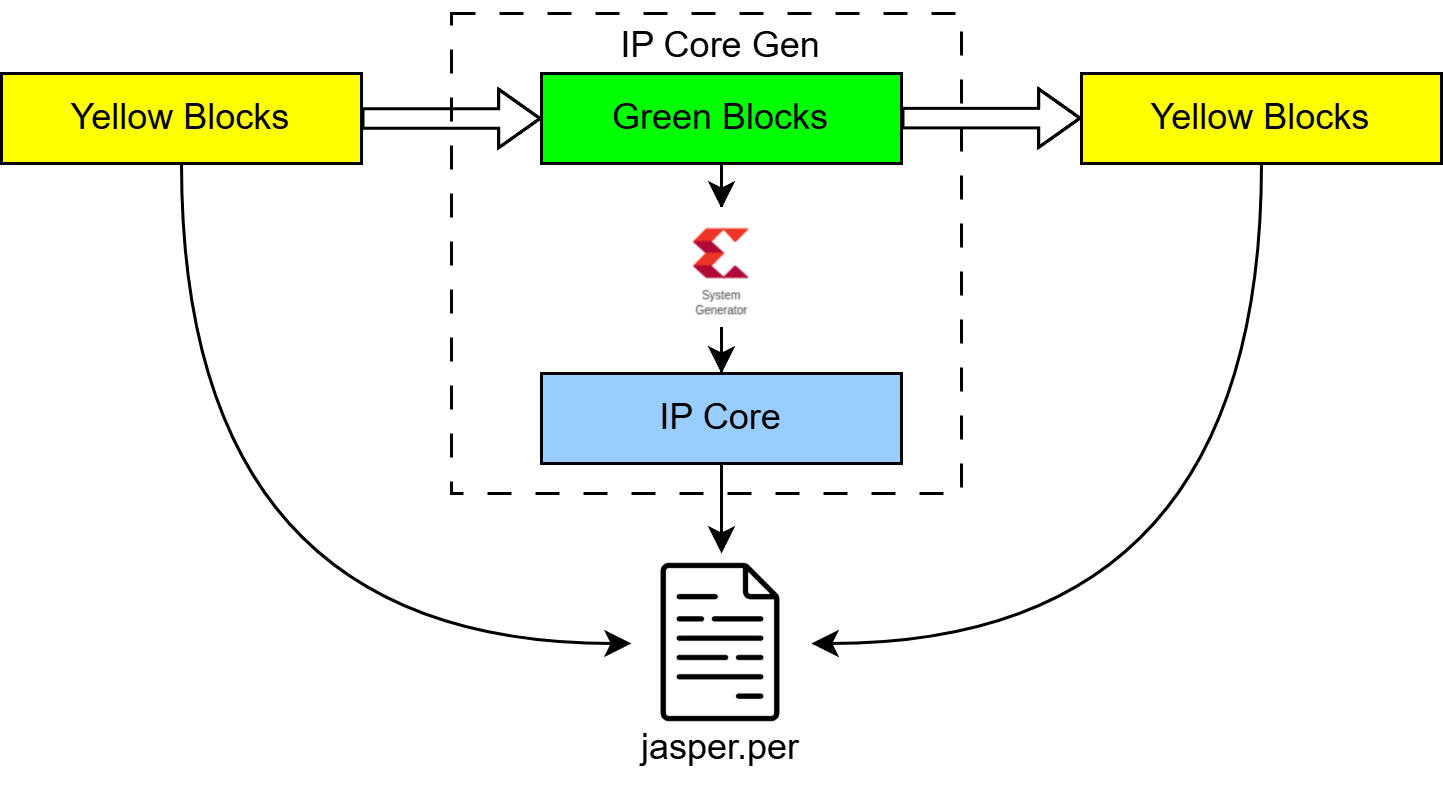}
    \caption{
        The generation of the $jasper.per$ file, which contains the configuration data of the Yellow blocks and the IP core information.
        \label{jasper_per_gen}
    }
\end{figure}

\indent Simulink combined with Model Composer offers a repository of AMD-specific blocks that enables designers to simulate, validate, and automatically create HDL code from Simulink models. This DSP IP core encompasses the configuration and link details of all DSP blocks and, coupled with the CASPER Yellow blocks, facilitates the interface to peripherals like ADCs, DACs, and other I/O components. The CASPER frontend gathers configuration data from the peripheral blocks and records this information in $jasper.per$ alongside the IP core details. The generation of the $jasper.per$ is shown in Figure~\ref{jasper_per_gen}. The file $jasper.per$ serves as the primary interface between the frontend and the backend.\\

\subsection{Backend}
\label{casper-backend}
\begin{figure}[htbp]
    \centering
    \includegraphics[height=8cm]{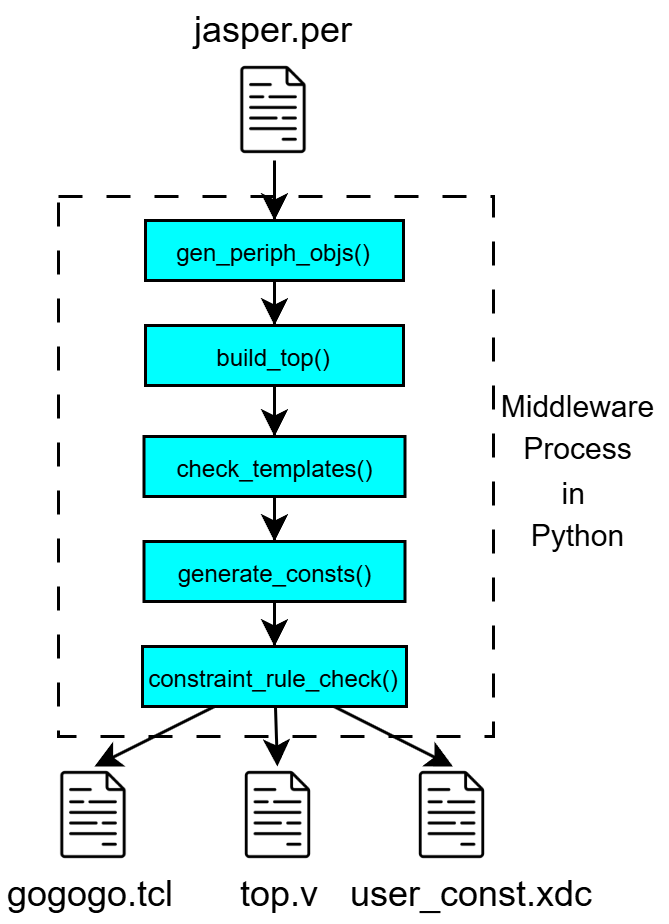}
    \caption{
        The main process of the generation of gogogo.tcl and top.v in Python:(1) Python scripts parse the $jasper.per$, and generate the peripherals objects based on the Yellow blocks information; (2) top module is created based on the peripherals objects; (3) Check for any Yellow blocks marked with a non-None value of the "template\_project" attribute; (4) Generate constraints for different Yellow blocks; (5) Check if the constraints are valid.}
    \label{gogogo_top_gen}
\end{figure}

\indent The $jasper.per$ file serves as the connector between the frontend and backend. The backend comprises two primary components: Python scripts and an FPGA design tool. The CASPER toolflow is primarily compatible with AMD Xilinx platforms, and the current FPGA design tool utilized is AMD Xilinx Vivado. The Python scripts within the CASPER toolflow complete the ``middleware process'', parsing the $jasper.per$ generated by the frontend and subsequently creating objects for each Yellow block. The Yellow block objects possess methods to incorporate HDL source code into a Vivado project by appending TCL commands to a file designated as $gogogo.tcl$, and to include HDL module instances within the top module titled $top.v$. In the middleware process, $user\_const.xdc$ is generated as a constraint file for the Vivado project, encompassing the pinout definition and time constraints. At the end of this phase, the TCL commands for synthesis and bitstream generation are incorporated into the TCL file. The generation of the $gogogo.tcl$, $top.v$ and $user\_const.xdc$ is shown in Figure~\ref{gogogo_top_gen}.\\

\begin{figure}[htbp]
    \centering
    \includegraphics[height=4cm]{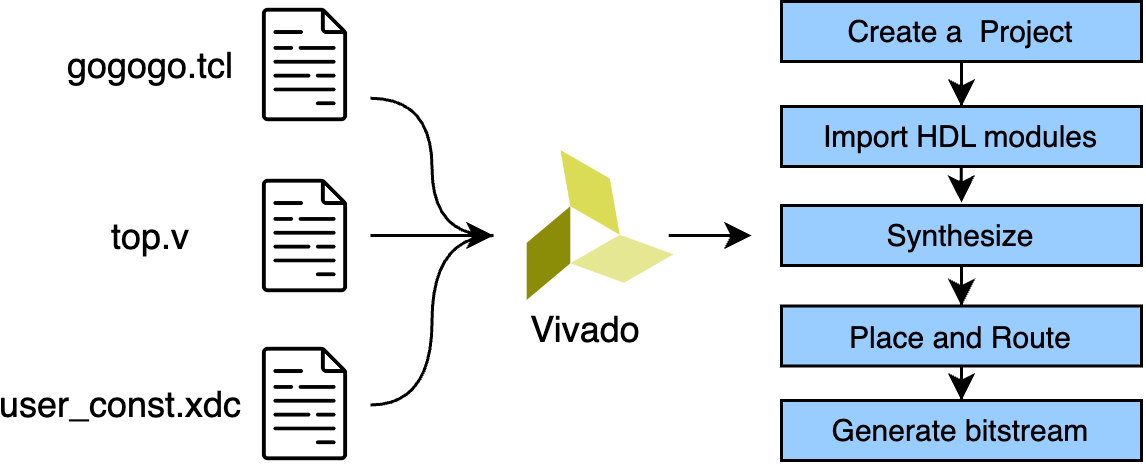}
    \caption{
        The workflow of Vivado for the bitstream generation.
        \label{vivado_workflow}
    }
\end{figure}

\indent After the generation of $gogogo.tcl$, $top.v$, and $user\_const.xdc$, Vivado will run in the background: A Vivado project is created utilizing the $gogogo.tcl$, importing all HDL files (including $top.v$), the IP core generated by the frontend, and the $user\_const.xdc$ into the Vivado project. Vivado then initiates the synthesis of the design, followed by placement and routeing, and finishing in bitstream generation. The Vivado working flow is shown in Figure~\ref{vivado_workflow}.\\

\begin{figure}[htbp]
    \centering
    \includegraphics[height=11.5cm]{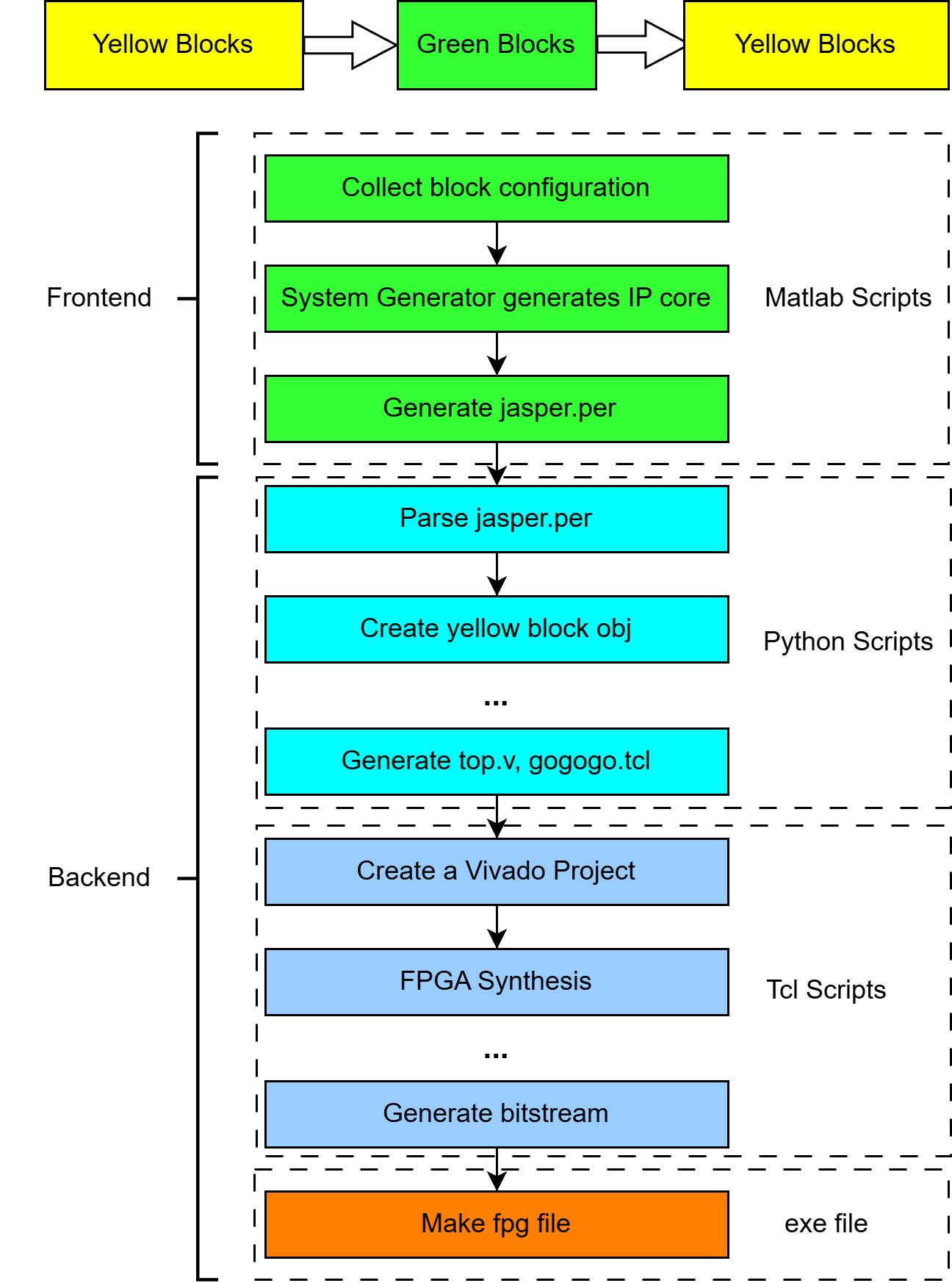}
    \caption{
        The workflow of the current CASPER toolflow.
        \label{casper_toolflow_workflow}
    }
\end{figure}

\indent Ultimately, a software called $mkfpg$ processes the created bitstream file and accompanying metadata to produce the final product file, which has a ".fpg" extension. The ".fpg" file will be uploaded to the FPGA platform, after which the FPGA will commence real-time signal processing. The workflow of the whole CASPER toolflow is shown in Figure~\ref{casper_toolflow_workflow}. \\

\section{CASPER toolflow in Scilab}
\label{casper_toolflow_in_scilab}
\indent Scilab is an open-source software platform intended for numerical analysis and scientific computation. It uses a similar syntax to MATLAB, and similarly allows the integration of additional features via toolboxes. This includes ``Xcos''\footnote{https://www.scilab.org/software/xcos}, a graphical editor for modeling and simulating in a similar manner to Simulink. Due to this and the similarity of the interface to that of Simulink to which CASPER users are already accustomed, Scilab was selected as an alternative graphical front-end option.\\

% The user interface of Scilab is shown in figure\ref{scilab_gui}, which is similar to MATLAB, which makes it easy for users to learn and use. This turns out Scilab is a good option to replace MATLAB in the CASPER toolflow.\\

% \begin{figure}[htbp]
%     \centering
%     \includegraphics[height=10cm]{figures/scilab_gui.png}
%     \caption{
%         The user interface of Scilab is similar to MATLAB.
%         \label{scilab_gui}
%     }
% \end{figure}

\begin{figure}[htbp]
    \centering
    \includegraphics[height=11.5cm]{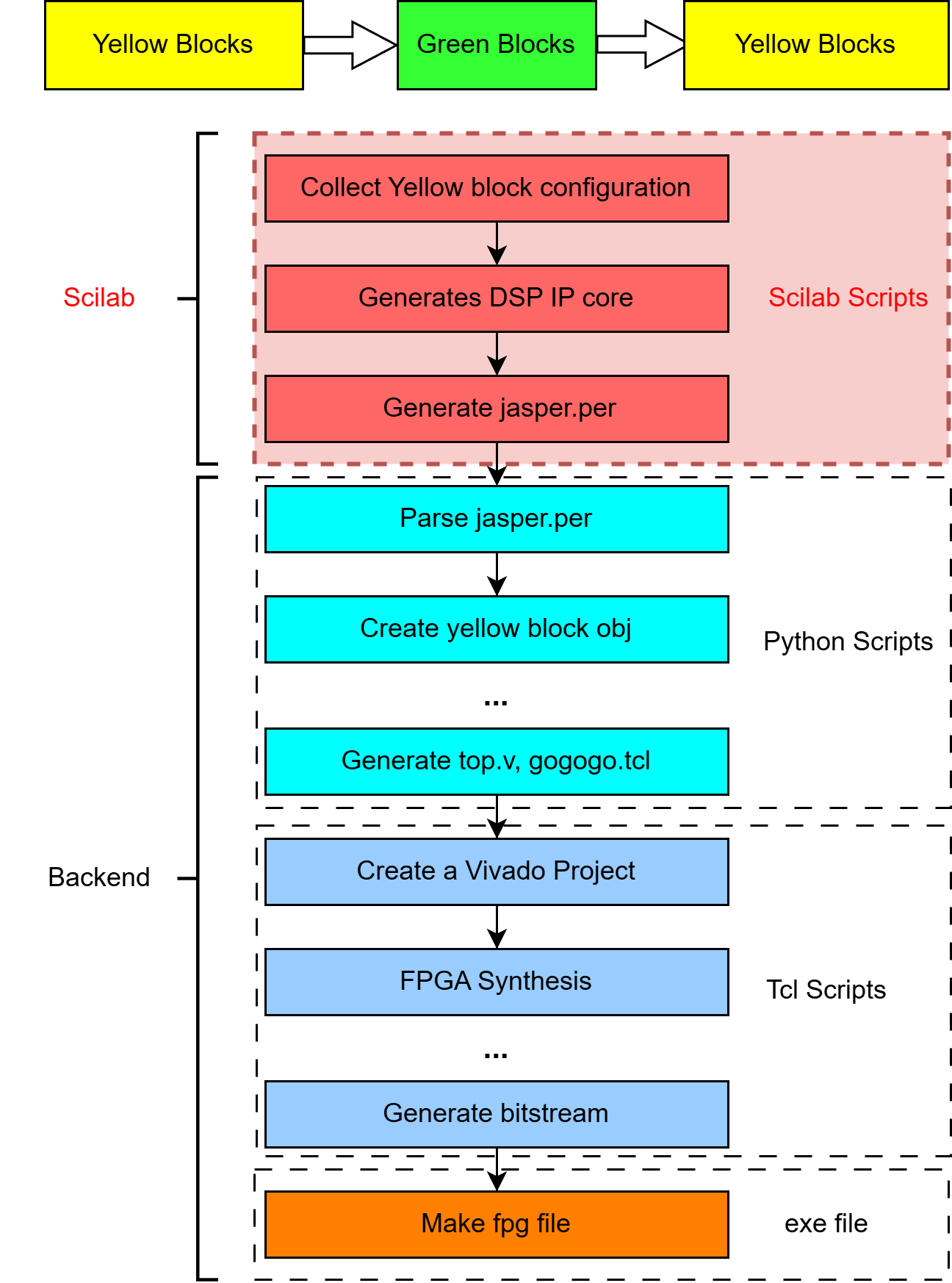}
    \caption{
        The workflow of the CASPER toolflow in Scilab. The important is step is collecting block information, generating DSP IP core and creating $jasper.per$.
        \label{scilab_toolflow_workflow}
    }
\end{figure}

\indent Figure~\ref{casper-toolflow-workflow} illustrates that $jasper.per$ is the important file connecting the frontend and backend in the CASPER toolflow. Consequently, the critical step in transitioning the frontend from MATLAB to Scilab is the generation of $jasper.per$ in Scilab. Since $jasper.per$ encompasses the details of the DSP IP core and the information of the Yellow blocks produced by Model Composer, it is essential to find out the processes for constructing Yellow/Green blocks, gathering the blocks' information, and generating the DSP IP core in Scilab. The workflow in Scilab is shown in Figure~\ref{scilab_toolflow_workflow}.\\

\newpage
\subsection{CASPER Blocks in Scilab}
\label{casper_blocks_in_scilab}
\indent Xcos is a graphical editor within Scilab that is utilized for modeling and simulation in a manner analogous to simulink in MATLAB. It is analogous in principle to Simulink in MATLAB. Each Xcos block specified in Xcos is represented as an object, encompassing a model data structure (scicos\_model\footnote{https://help.scilab.org/scicos\_model.html}), a graphical data structure (scicos\_graphics\footnote{https://help.scilab.org/scicos\_graphics.html}), and GUI data. The model data structure includes such information as the number of ports, the ports id and the bitwidth of each port. The graphic data structure contains the visual characteristics of the block. \\
\indent CASPER Yellow/Green blocks are defined in Xcos, represented as $.sci$ files. However, in the case of Xcos, both the Yellow and Green blocks serve as a wrapper for an HDL module. This wrapper is required to provide the block information to the toolflow. The required data structure is given in Table~\ref{casper_data_struct_gen}. \\
\begin{table}[h]
    \centering
    \begin{tabular}{|c|c|c|}
        \hline
        CASPER Block       & Source              \\
        \hline
        Block Type         & model.label         \\
        Block Tag          & gui                 \\
        Block Name         & graphics.exprs      \\
        Parameters         & graphics.exprs      \\
        In/Out Port Name   & model.in/out\_label \\
        In/Out Port Number & model.in/out        \\
        In/Out Port Width  & model.in2/out2      \\
        \hline
    \end{tabular}
    \caption{The data source for generating the CASPER block data structure.}
    \label{casper_data_struct_gen}
\end{table}

\begin{figure}[htbp]
    \centering
    \includegraphics[height=5cm]{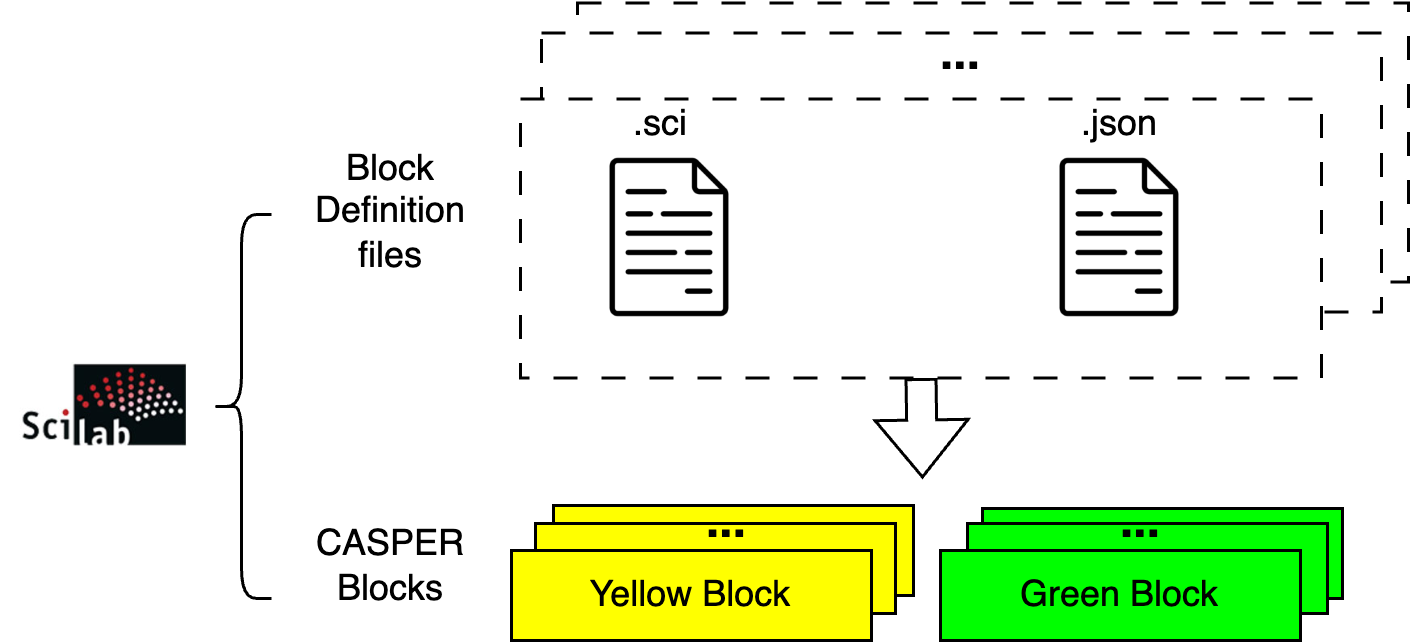}
    \caption{
        A CASPER block in Xcos is defined by a $.sci$ file and a $.json$ file. The $.sci$ file defines the GUI of the block, while the $.json$ contains the parameters set by users.
        \label{xcos_block_definiation}
    }
\end{figure}

\indent The block GUI information is outlined in the $.sci$ file. As users configure these blocks to suit the application, the parameters are not hard-coded, but defined in a $.json$ file to allow for modification. A CASPER Yellow/Green block in Xcos requires both a $.sci$ file and a $.json$ file definition, as illustrated in Figure~\ref{xcos_block_definiation}. An example of the block definition files can be found here\footnote{https://github.com/casper-astro/mlib\_devel/blob/scilab-m2021a/scilab\_library/scilab\_blocks/casper\_xps/swreg.sci}\footnote{https://github.com/casper-astro/mlib\_devel/blob/scilab-m2021a/scilab\_library/scilab\_blocks/casper\_xps/swreg.json}.\\

\begin{figure}[htbp]
    \centering
    \subfloat[Software register block in Simulink]{\includegraphics[height=7cm]{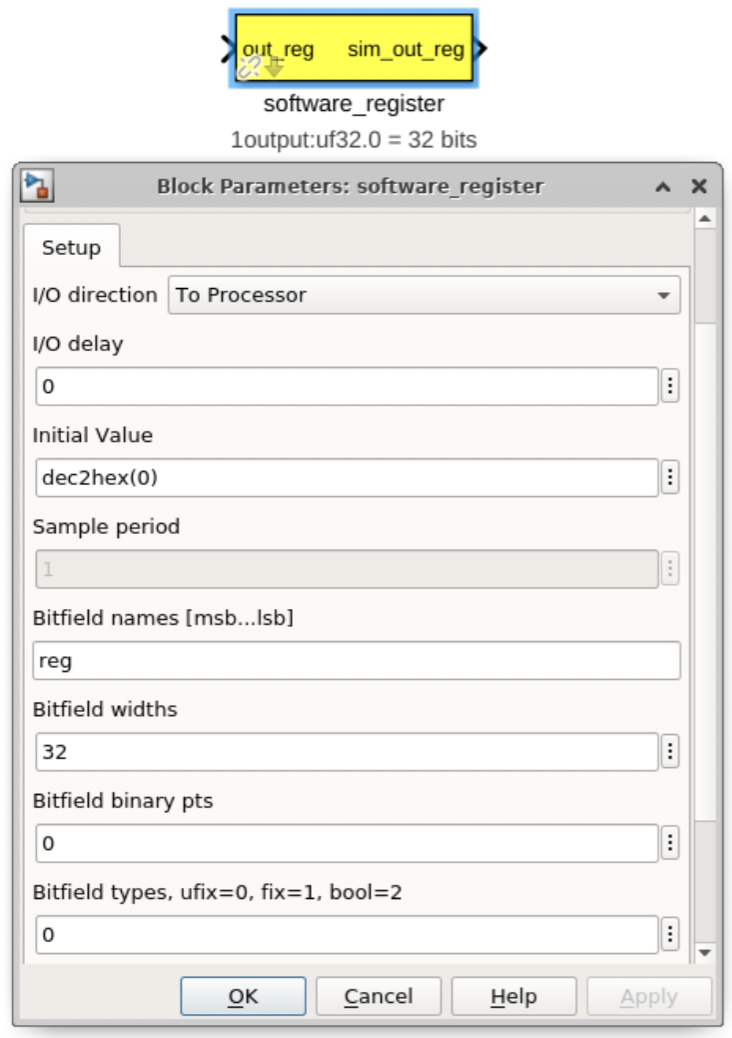}}\hspace{5pt}
    \subfloat[Software register block in Xcos]{\includegraphics[height=7cm]{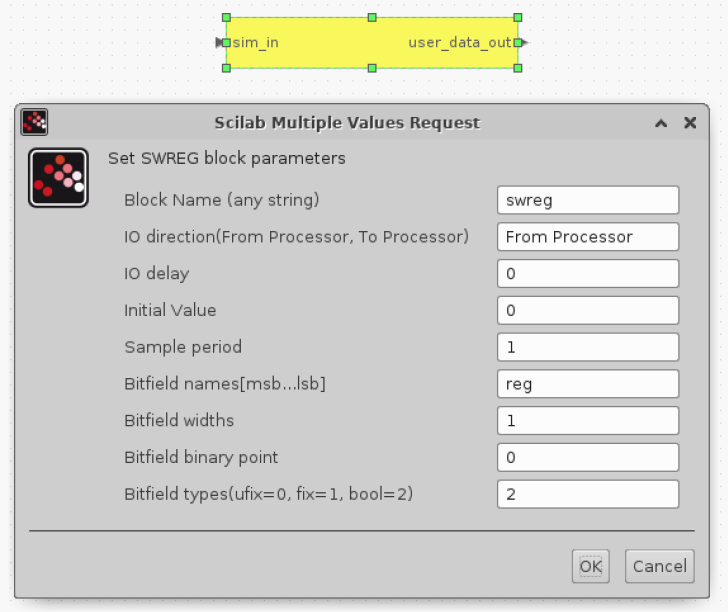}}\\
    \caption{A comparison between software register blocks in Xcos and in Simulink. The GUIs look similar, so that it will be easy to use the new frontend.}
    \label{block_comparison_in_scilab_matlab}
\end{figure}

\indent The CASPER toolflow is now approximately 20 years old. As the interface has remained mostly the same during this time period, it is useful from a user perspective to keep the GUI for each block interface similar to the Simulink original. A comparison of blocks in Xcos and Simulink is presented in Figure~\ref{block_comparison_in_scilab_matlab}.\\

\subsection{Block Information Collection}
\begin{figure}[htbp]
    \centering
    \includegraphics[height=4cm]{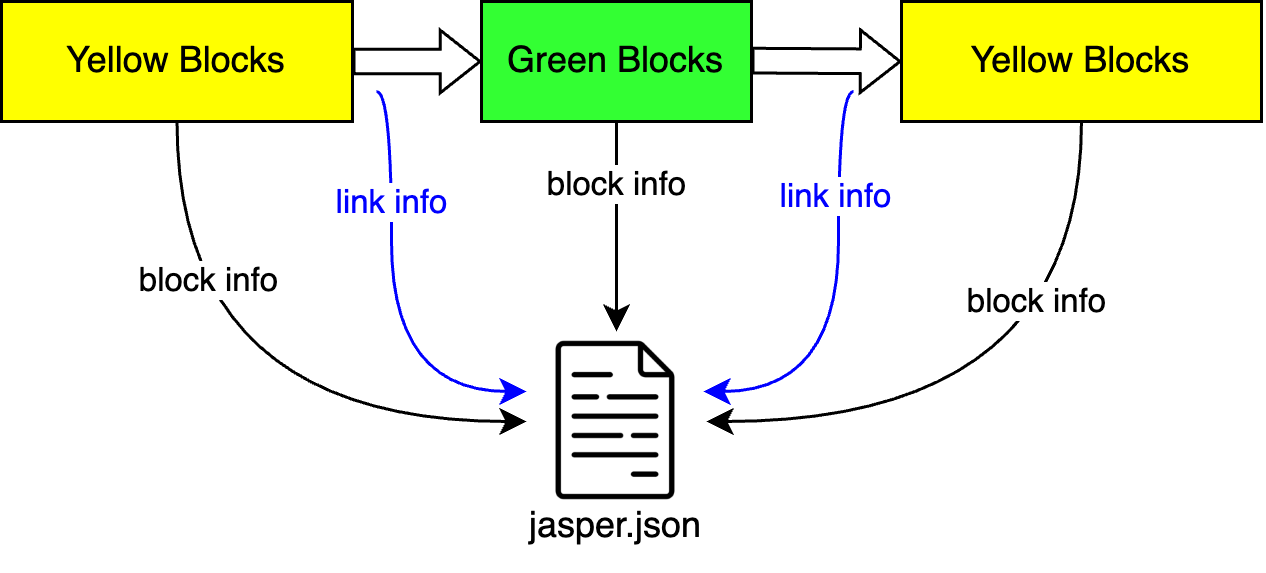}
    \caption{
        All of the necessary information are collected, including block name, block parameters, block ports data and link information. These information are stored in $jasper.json$.
        \label{block_information_collect}
    }
\end{figure}

\indent As stated in Section~\ref{casper_blocks_in_scilab}, all requisite block information is contained within the scicos\_model and scicos\_graphics data structures. Examining all block objects' scicos\_model and scicos\_graphics collects all Yellow/Green block information, encompassing block name, parameters, and port data. In addition to block information, link information is also essential, indicating the connections between blocks. In Simulink, AMD Model Composer autonomously manages the link information. Xcos lacks a tool comparable to AMD Model Composer, so it is necessary to gather the link information for the subsequent stage. In Scilab, all relevant information, including block and link data, is collected and stored in a pre-middleware file designated as $jasper.json$. The creation of $jasper.json$ is illustrated in Figure~\ref{block_information_collect}.\\

\subsection{DSP IP Core Generation}
\begin{figure}[htbp]
    \centering
    \includegraphics[height=2.5cm]{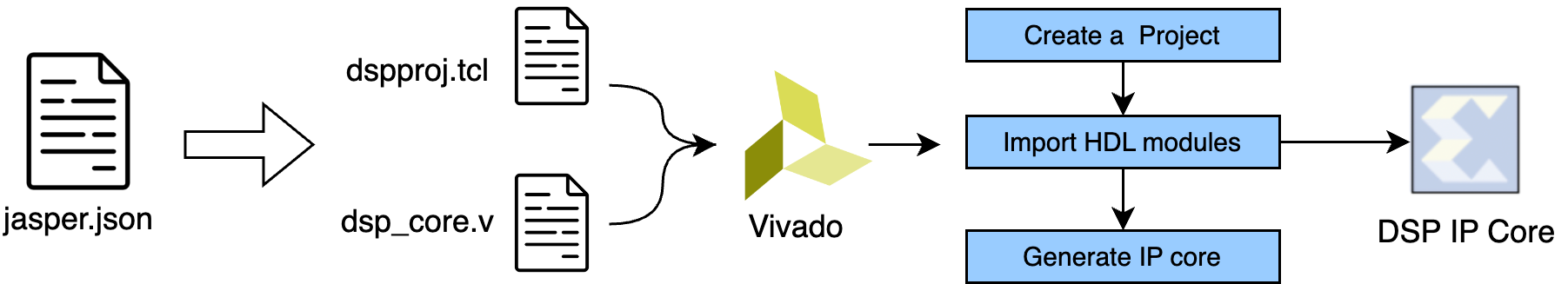}
    \caption{
        The DSP blocks' info is collected from $jasper.json$ first, then a top module ($dsp\_core.v$) for the DSP IP core is created automatically. After that, a TCL script is created automatically, importing all of the DSP modules and the top module to a Vivado project, and creating the DSP IP core automatically.
        \label{scilab_ip_gen}
    }
\end{figure}

As indicated in Section~\ref{casper-frontend}, AMD Model Composer autonomously generates the DSP IP core. Generating the similar DSP IP core in Scilab is essential for utilizing the same backend. A Vivado Tcl script ($dspproj.tcl$) can be automatically generated from the block information in the $jasper.json$, allowing the import of the DSP HDL modules into a Vivado project. Utilizing the link information in the $jasper.json$, a primary module ($dsp\_core.v$) of the DSP IP core can then be generated automatically. This automatic generation of the IP core is performed by Vivado in the background. The creation of this IP core is illustrated in Figure~\ref{scilab_ip_gen}.\\

\subsection{Generate $jasper.per$}
\begin{figure}[htbp]
    \centering
    \includegraphics[height=4.5cm]{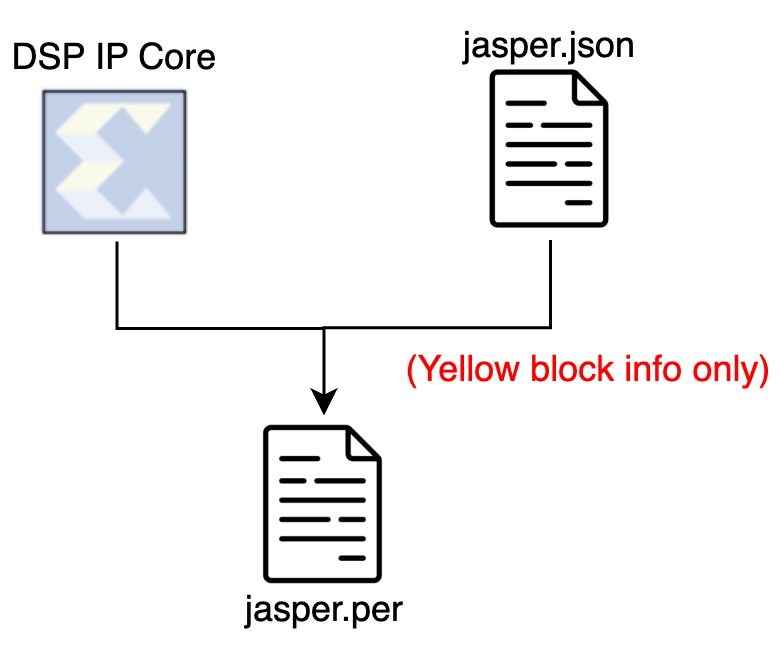}
    \caption{
        $jasper.per$ is generated based on the $jasper.json$ and the generated DSP IP core. Yellow block information in the $jasper.json$ is required only for the $jasper.per$.
        \label{scilab_jasper_per_gen}
    }
\end{figure}

\indent The $jasper.per$ can be automatically constructed by parsing the $jasper.json$ to obtain the path of the generated DSP IP core. The $jasper.json$ file encompasses all information, including that of Yellow and Green blocks, whereas the $jasper.per$ file contains solely the information pertaining to Yellow blocks. As different keys (e.g., xps\_blocks, dsp\_blocks) are used for different types of blocks in $jasper.json$,  it's straightforward to parse the $jasper.json$ to extract information solely pertaining to the Yellow blocks. The production of $jasper.per$ is illustrated in Figure~\ref{scilab_jasper_per_gen}.\\

\subsection{Simulation in Scilab}
\begin{figure}[htbp]
    \centering
    \includegraphics[height=5.3cm]{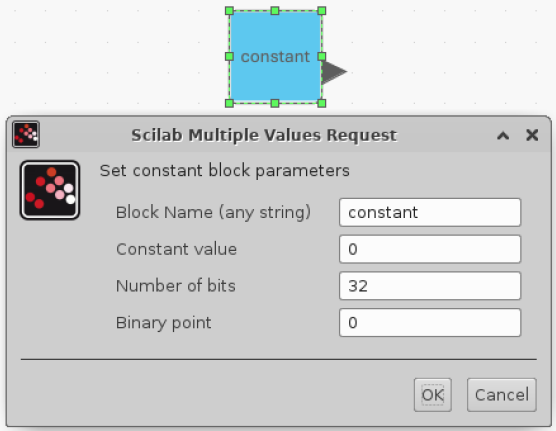}
    \caption{
        An example of a blue block, constant block. It generates constant data for the CASPER design simulation in Xcos.
        \label{blue_block_constant}
    }
\end{figure}

\indent The simulation in Simulink is primarily for Green blocks (DSP blocks). Yellow blocks are peripheral and so excluded from the simulation, or used simply as placeholder for source generation within Simulink. The simulations are bit and clock true. To create a similarly accurate simulation in Scilab, a Vivado project has been established encompassing all DSP blocks in the Scilab model. A testbench and corresponding input data are necessary to execute the simulation. The input data can be generated by getting the simulation blocks information from $jasper.json$. Then the data file will be opened in the testbench, and the data will be read out as the input simulation data in the testbench. To produce input data for the testbench, a new collection of blocks is established in Xcos - Blue blocks. The Blue blocks comprise two types: Blue source blocks and Blue sink blocks. Blue source blocks, such as sine wave blocks, supply the essential parameter for the incoming data. Python scripts gather these parameters and produce the input data for the testbench. Figure~\ref{blue_block_constant} illustrates an example of a blue block. So far, we have sine wave block and constant block for simulation. More simulation blocks (like white noise block and others) will be added in the future.\\

\begin{figure}[htbp]
    \centering
    \includegraphics[height=3.5cm]{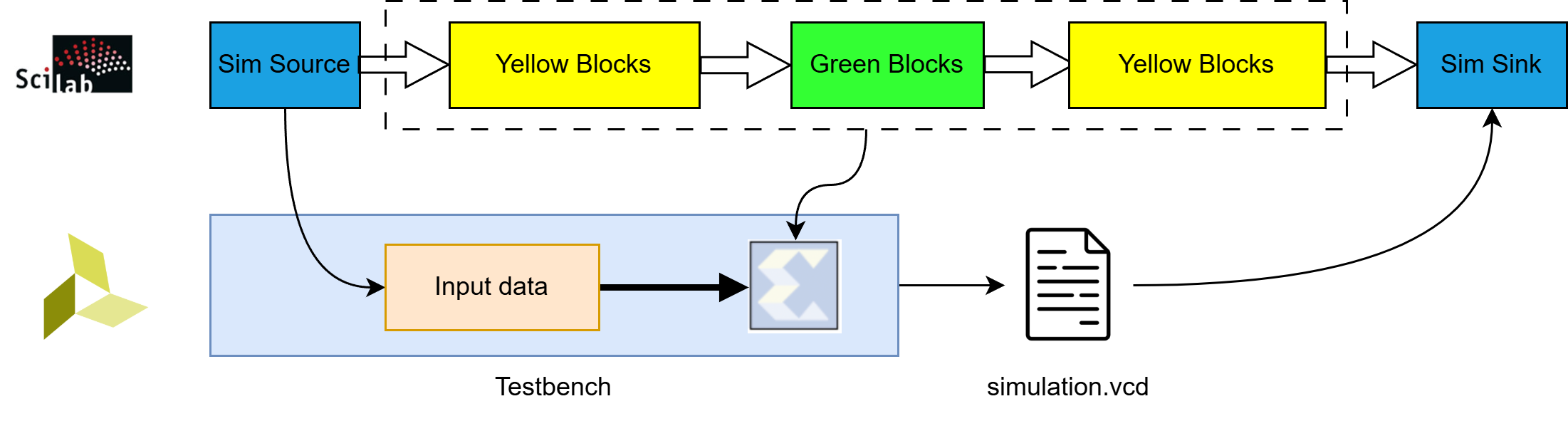}
    \caption{
        The simulation flow in Scilab: Blue source blocks generate input data for the testbench, which is generated automatically for the DSP IP core. Vivado runs the simulation in the background, and exports the simulation results in a VCD file. Blue sink blocks get the simulation data from the VCD file, and show the results visually. The timing validation is implemented by the Python scripts, and is done by the backend.}
    \label{scilab_simulation}
\end{figure}

\indent The Vivado simulator is utilized for simulation and operates in the background. Upon completion of the simulation, the results are exported and stored in a VCD\footnote{https://en.wikipedia.org/wiki/Value\_change\_dump} file. The Blue sink block retrieves simulation data from the Value Change Dump (VCD) file and displays the results visually. The simulation process is illustrated in Figure~\ref{scilab_simulation}. The source code of the new frontend for the toolflow is here\footnote{https://github.com/casper-astro/mlib\_devel/tree/scilab-m2021a}.\\

\begin{figure}[htbp]
    \centering
    \includegraphics[height=4cm]{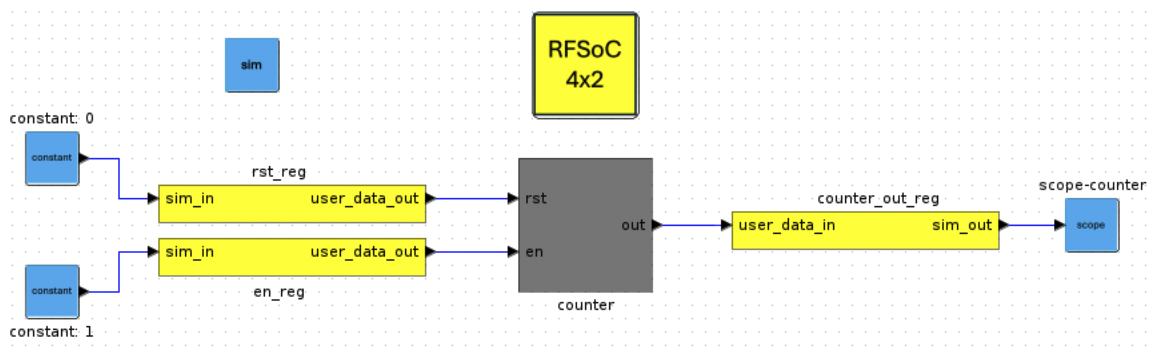}
    \caption{
        A counter example with two constant simulation blocks and one scope simulations.
        \label{counter_sim}
    }
\end{figure}

\begin{figure}[htbp]
    \centering
    \includegraphics[height=3cm]{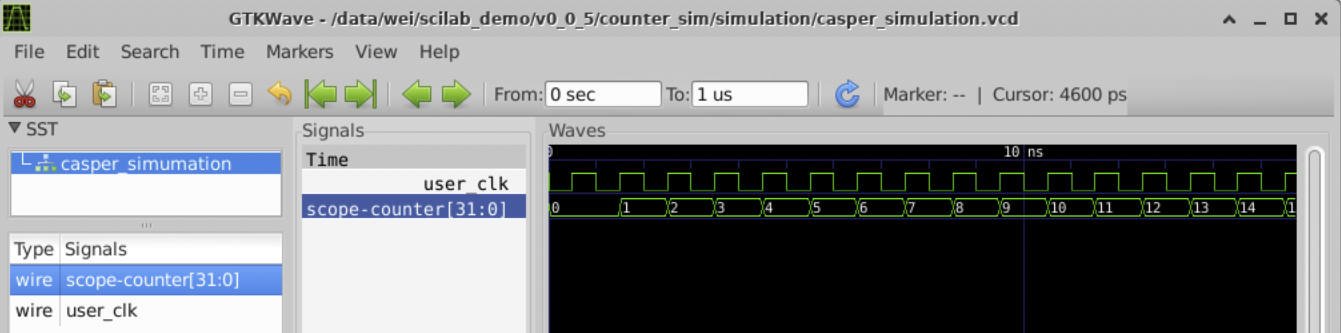}
    \caption{
        The simulation results are shown in GTKWave.
        \label{gtkwave_sim}
    }
\end{figure}

\begin{figure}[htbp]
    \centering
    \includegraphics[height=6.5cm]{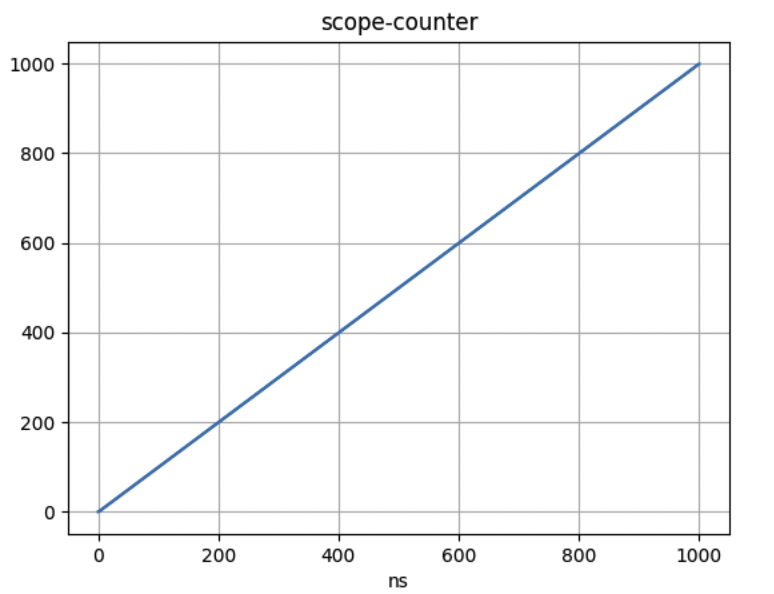}
    \caption{
        The simulation results are plotted with Matplotlib.
        \label{pyplot_sim}
    }
\end{figure}

\indent As the simulation results are recorded in a VCD file, users can use any visualization tools to show the results. In the current toolflow, Matplotlib\footnote{https://matplotlib.org/} and GTKWave\footnote{https://gtkwave.sourceforge.net/} are the two options for showing the simulation results. A counter example is shown in Figure~\ref{counter_sim}. In the example, the scope block is incorporated into the output port; however, scopes can be integrated at any location within the design, as the simulation results of all signals are contained in the VCD file. (An example of multiple scopes within the design is shown in the spectrometer demonstration outlined in Figure~\ref{spectrometer_demo_design} in Section~\ref{spectrometer-demo}.) The counter simulation results are shown Figure~\ref{gtkwave_sim} and Figure~\ref{pyplot_sim}, which demonstrate the simulation flow works.\\

\begin{figure}[htbp]
    \centering
    \includegraphics[height=5cm]{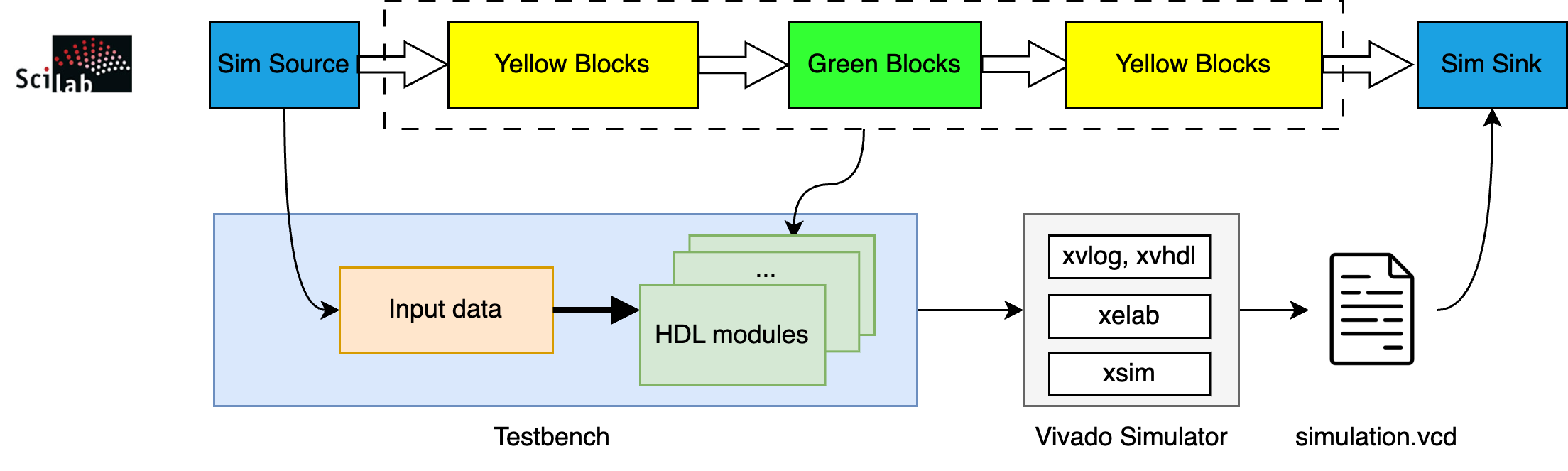}
    \caption{
        An alternative simulation flow: Blue source blocks generate input data for the testbench. Then the source file of the testbench and all of the used HDL modules are compiled, elaborated and linked, and simulated by Vivado simulator directly, which skips the IP core generation in the previous simulation method. (xvlog, xvhdl, xelab and xsim is a part of Vivado simulator.)}
    \label{new_simulation_method}
\end{figure}

\indent The above method has been implemented in the current version of the toolflow. While the simulation itself is of a comparable speed to directly simulating the HDL in Vivado, the necessity of launching a Vivado instance and generating the IP core  creates an initial delay. To speed up the simulation time, another simulation method is explored, which skips the IP core generation, and runs the simulation directly based on the HDL modules. The diagram of the this simulation flow is illustrated in Figure~\ref{new_simulation_method}.

\indent A simulation speed comparison between Scilab frontend and Simulink frontend is based on the spectrometer design shown in Figure~\ref{spectrometer_demo_design} and Figure~\ref{spectrometer_demo_design_simulink} in Section~\ref{spectrometer-demo}. When running simulation with the IP core generation, it takes 1.5~min(generate IP core) + 1~min (run simulation) to finish the simulation in Scilab, while it takes $\sim$20 seconds to finish the simulation in Simulink. When using the direct simulation method, which is based on each individual HDL source file, it takes $\sim$27 seconds to finish the simulation. Future work will seek to identify and reduce any remaining overheads.\\

\newpage
\subsection{Adaptable Architecture}
\label{other-frontend}
\begin{figure}[htbp]
    \centering
    \includegraphics[height=8cm]{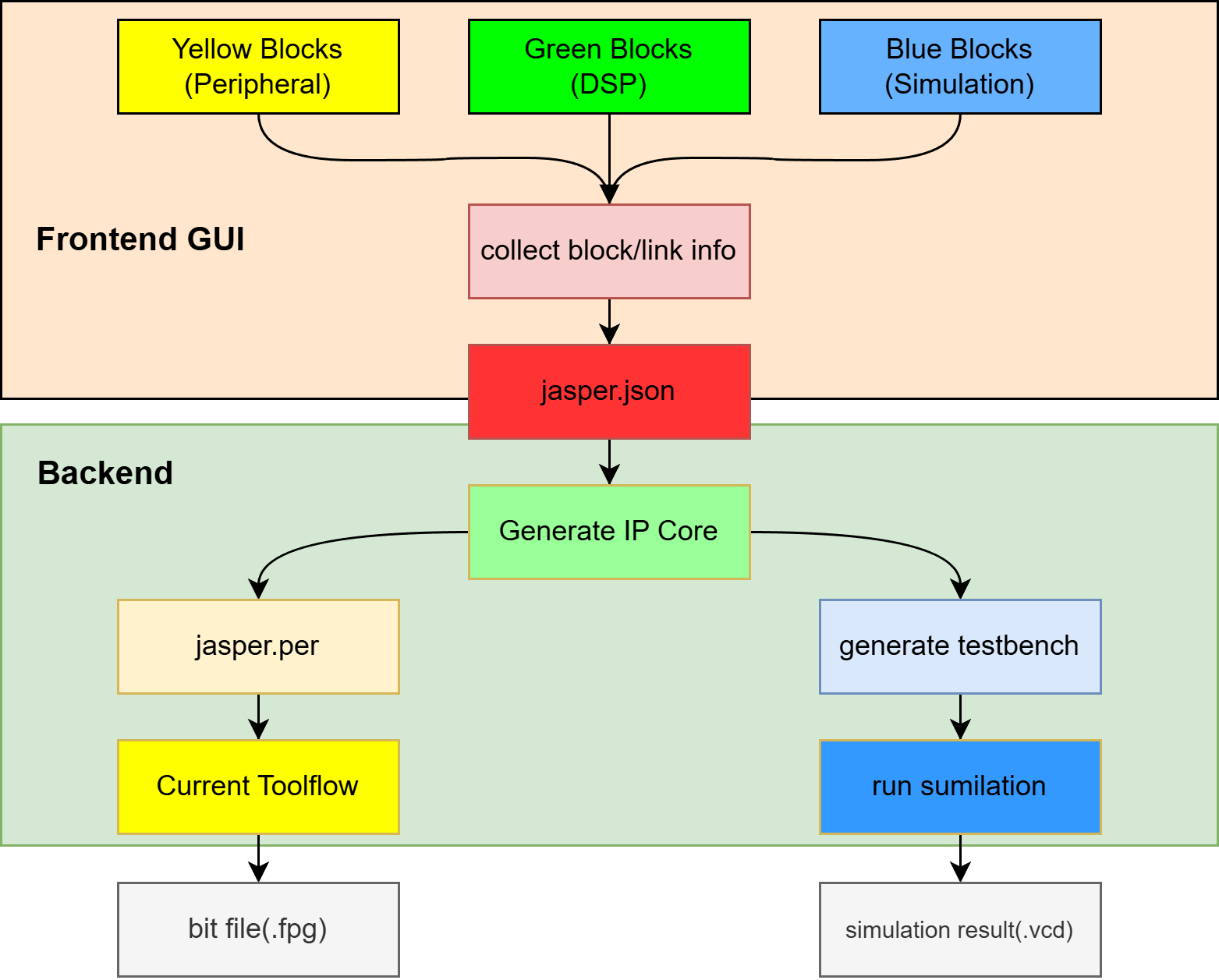}
    \caption{
        The architecture of the new CASPER toolflow, which consists of frontend GUI and the backend.
        \label{casper_arch}
    }
\end{figure}

\indent Even though the target of this work is to develop the toolflow interfacing with Scilab, the architecture was designed to allow the use of alternative frontend tools with minimal effort. This architecture is summarized in Figure~\ref{casper_arch}. The frontend GUI offers users Yellow, Green, and Blue blocks, while the scripts executed within the frontend GUI gather block and link information for the creation of $jasper.json$. The new backend, comprising Python scripts and FPGA synthesis tools, operates in the background, and assumes control of other tasks. The adaptable design facilitates the porting of the CASPER toolflow to other open-source frontends, such as GNU Radio\footnote{https://github.com/casper-astro/gr-casper}, should users choose to do so. At his stage, however, CASPER is focused solely on using Scilab in this role.\\

% \indent  For example purposes, a straightforward demonstration of the CASPER toolflow functioning within GNU Radio is illustrated in Figure~\ref{gnuradio_demo}. This demonstration utilizes an RFSoC4x2 block, an adder block, and three software register blocks. The generated "$.fpg$" file has been validated on the RFSoC4x2 platform to ensure the design functions correctly. The GNU Radio workflow and the blocks used in this demonstration are available here\\

% \begin{figure}[htbp]
%     \centering
%     \includegraphics[height=6.5cm]{figures/gnuradio_demo.png}
%     \caption{
%         A demonstration of a design for a simple adder based on CASPER blocks in GNU Radio.
%         \label{gnuradio_demo}
%     }
% \end{figure}

% \begin{figure}[htbp]
%     \centering
%     \includegraphics[height=3cm]{figures/gnuradio_result.png}
%     \caption{
%         The compiled result for the demo in GNURadio.
%         \label{gnuradio_result}
%     }
% \end{figure}

\clearpage
\newpage
\section{Spectrometer Demonstration based on CASPER Toolflow in Scilab}
\label{spectrometer-demo}
\subsection{Spectrometer Demonstration in Scilab}
\begin{figure}[htbp]
    \centering
    \includegraphics[height=2.5cm]{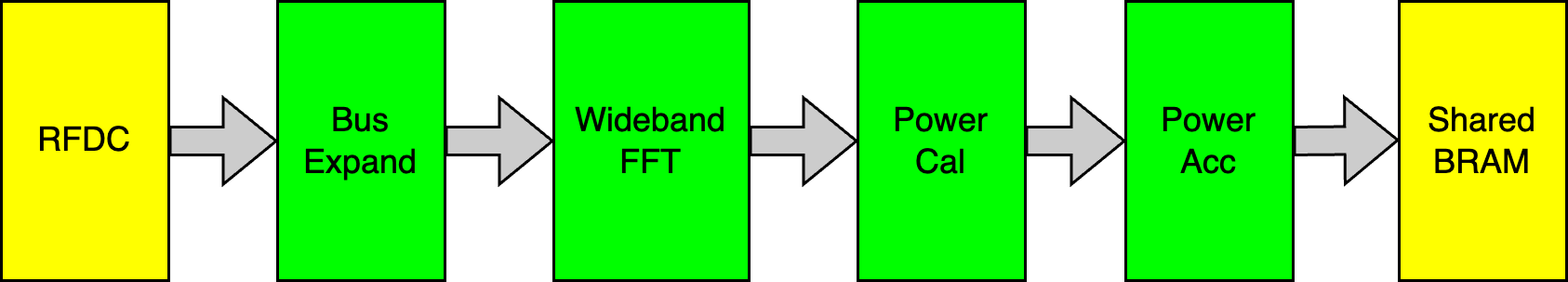}
    \caption{
        The block diagram of the spectrometer demonstration.}
    \label{spectrometer_demo_block}
\end{figure}

\begin{figure}[htbp]
    \centering
    \includegraphics[height=9cm,angle=90]{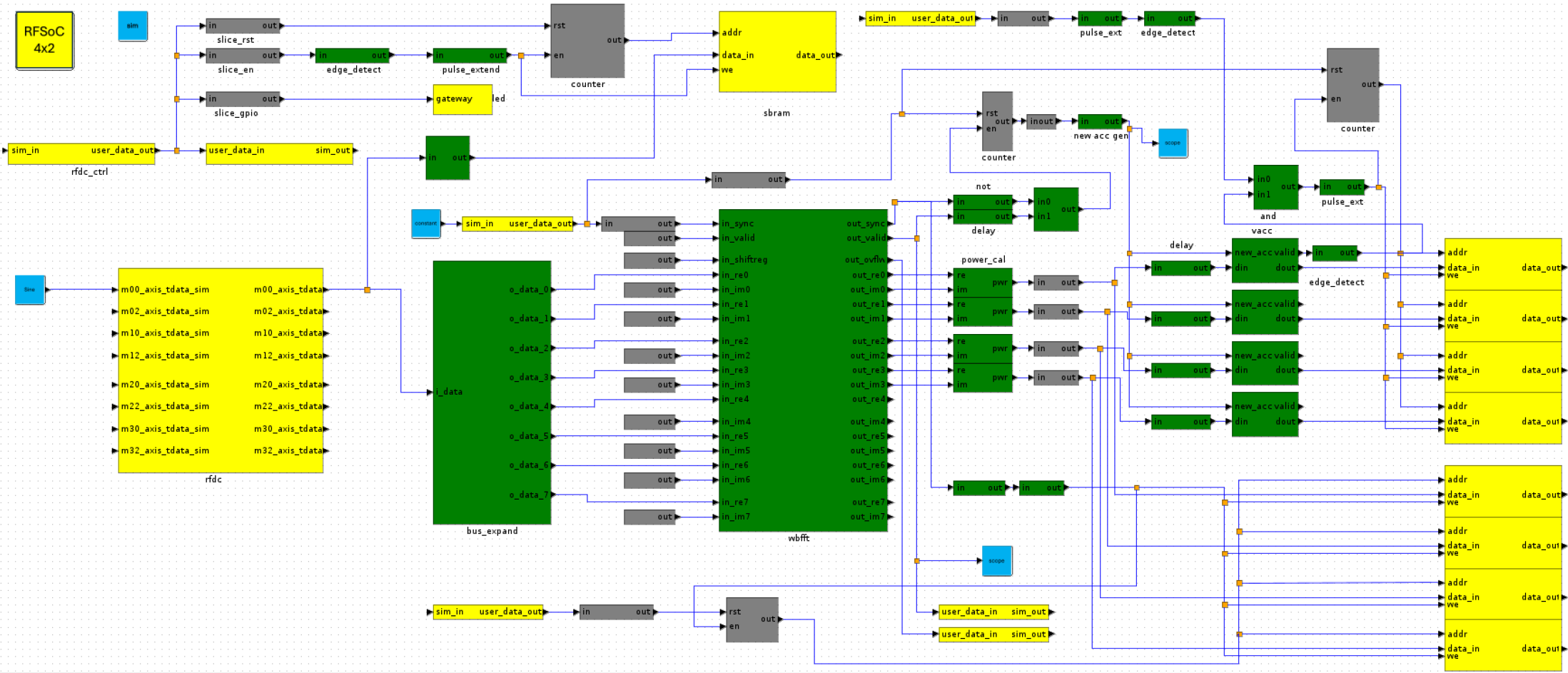}
    \caption{
        A spectrometer demonstration on RFSoC4x2 platform in Scilab, which measures its power as a function of frequency, and produces a spectrum. The RFDC block converts the analog signal to digital data, and then send the digital data to a Wideband FFT block. The Wideband FFT block finishes the channelization processing, and four simple\_bram\_vacc blocks accumulate the spectra data. The accumulated spectra data is stored in four Shared BRAM blocks.
    \label{spectrometer_demo_design}
    }
\end{figure}

\indent A spectrometer demonstration has been developed utilizing Yellow and Green blocks in Scilab. The design operates on an RFSoC4x2 platform. The primary components in this demonstration include an RFDC Yellow block, a Wideband FFT Green block, several Accumulation blocks, and several shared BRAM blocks. The underlying CASPER HDL implementations for the Wideband FFT and Accumulation blocks that were converted to work in Scilab for this example are provided on github\footnote{\label{casper_hdl}https://github.com/casper-astro/casper\_dspdevel}. Several blues blocks have been incorporated into the demonstration design for simulation purposes. The block diagram of the spectrometer is shown in Figure~\ref{spectrometer_demo_block}, and the demonstration design is illustrated in Figure~\ref{spectrometer_demo_design}. The key specifications are presented in Table~\ref{spectrometer_demo_block_settings}. \\

% The configurations of the essential blocks are presented in Table~\ref{spectrometer_demo_block_settings}.\\

\begin{table}[htbp]
    \centering
    \begin{tabular}{|l|llll|}
        \hline
        Specifications      & \multicolumn{4}{l|}{Values}          \\ \hline
        Sampling Rate       & \multicolumn{4}{l|}{1966.08 MSps}    \\ \hline
        Bandwidth           & \multicolumn{4}{l|}{983.04 MHz}      \\ \hline
        Spectral Channels   & \multicolumn{4}{l|}{1024}            \\ \hline
        Spectral Resolution & \multicolumn{4}{l|}{1.92 MHz}        \\ \hline
        Accumulation Length & \multicolumn{4}{l|}{8}               \\ \hline
        Integration Time    & \multicolumn{4}{l|}{$\sim$~4~$\mu$s} \\ \hline
    \end{tabular}
    \caption{The key specifications of the spectrometer demonstration.}
    \label{spectrometer_demo_block_settings}
\end{table}

% \begin{table}[htbp]
%     \centering
%     \begin{tabular}{|l|l|l|}
%         \hline
%         Blocks                              & Parameters                       & Values   \\ \hline
%         \multirow{4}{*}{Wideband FFT}       & Input Data Width                 & 16       \\ \cline{2-3}
%                                             & Output Data Width                & 16       \\ \cline{2-3}
%                                             & Number of Stream                 & 8        \\ \cline{2-3}
%                                             & Number of FFT Points             & 1024     \\ \hline
%         \multirow{4}{*}{Simple\_bram\_vacc} & Vector Length                    & 128      \\ \cline{2-3}
%                                             & Input Data Width                 & 32       \\ \cline{2-3}
%                                             & Output Data Width                & 32       \\ \cline{2-3}
%                                             & Data Type                        & UNSIGNED \\ \hline
%         \multirow{2}{*}{Shared BRAM}        & Addr Width(2\textasciicircum{}x) & 7        \\ \cline{2-3}
%                                             & Bitwidth                         & 32       \\ \hline
%     \end{tabular}
%     \caption{\textcolor{red}{The settings of the wideband FFT block (Wideband FFT), vector accumulation block (simple\_bram\_vacc) and shared Block RAM blocks (Shared BRAM). The accumulation signal is issued every 1024 clocks, so every 8 spectra are accumulated.}}
%     \label{spectrometer_demo_block_settings}
% \end{table}

\begin{figure}[htbp]
    \centering
    \includegraphics[height=8cm,angle=90]{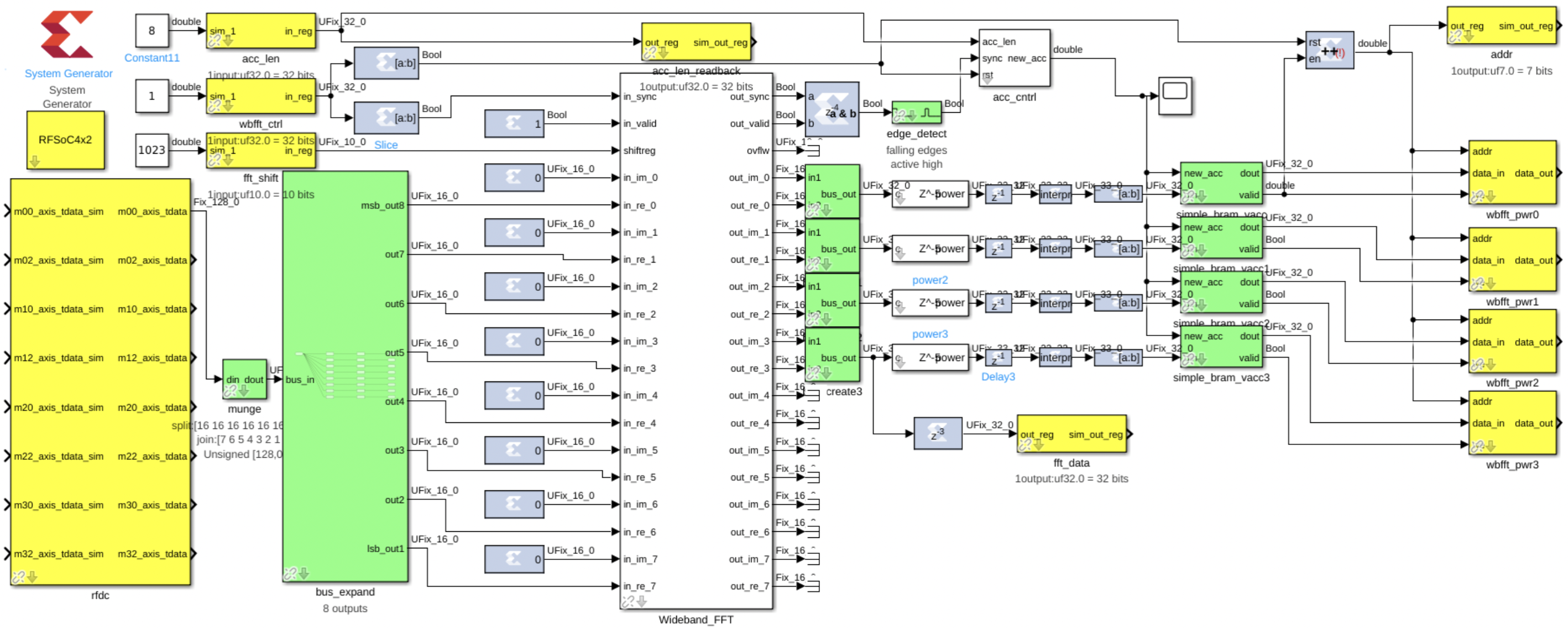}
    \caption{
        A spectrometer demonstration on RFSoC4x2 platform in Simulink for comparison with the demonstration in Scilab. This demonstration design contains the same core blocks, like RFDC block, Wideband FFT block, Accumulation blocks and Shared BRAM blocks.}
    \label{spectrometer_demo_design_simulink}
\end{figure}

\indent To validate the function of the CASPER blocks in Scilab, the same spectrometer is also implemented in Simulink with the same key blocks, like the RFDC block, Wideband FFT block, Accumulation blocks and Share BRAM blocks. The Simulink version of the same CASPER HDL blocks\footref{casper_hdl} were used. The demonstration design in Simulink is illustrated in Fig~\ref{spectrometer_demo_design_simulink}. A comparison between the spectrometer demonstration designs in Scilab and Simulink will be shown in Section~\ref{on_board_test}.\\

\clearpage
\newpage
\subsection{On-board Test}
\begin{figure}[htbp]
    \centering
    \includegraphics[height=3cm]{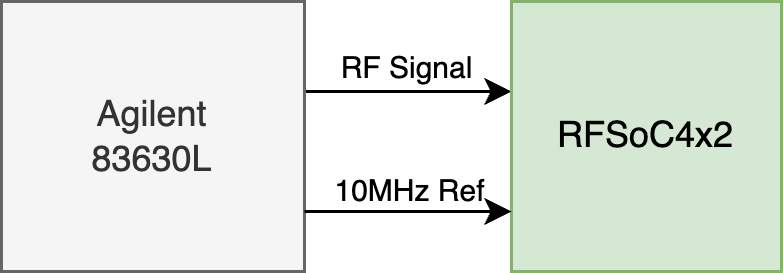}
    \caption{
        The test setup for the spectrometer demonstration design. It includes an Agilent synthesized swept-CW generator and a RFSoC4x2 board. A reference clock is provided from the Agilent signal generator to the RFSoc4X2 platform for synchronization.
        \label{test_setup}
    }
\end{figure}

\indent The testing configuration is depicted in Figure~\ref{test_setup}. This comprises an Agilent synthesized swept-CW generator\footnote{https://www.keysight.com/us/en/product/83630L/synthesized-sweptcw-generator-10-mhz-to-265-ghz.html} and a RFSoC4x2 board. One ADC channel is utilized for this test, with a sample frequency established at 1966.08 MSps. The input signal frequency to the RFSoC is 51.84MHz, with a power level of -15 dBm. The Agilent signal generator outputs a 10MHz reference signal to the RFSoC4x2 board, to get the RFSoC4x2 synchronous to the signal generator.\\
\label{on_board_test}

\begin{figure}[htbp]
    \centering
    \subfloat[The spectra result from the demonstration design in Scilab.]{
        \includegraphics[height=6cm]{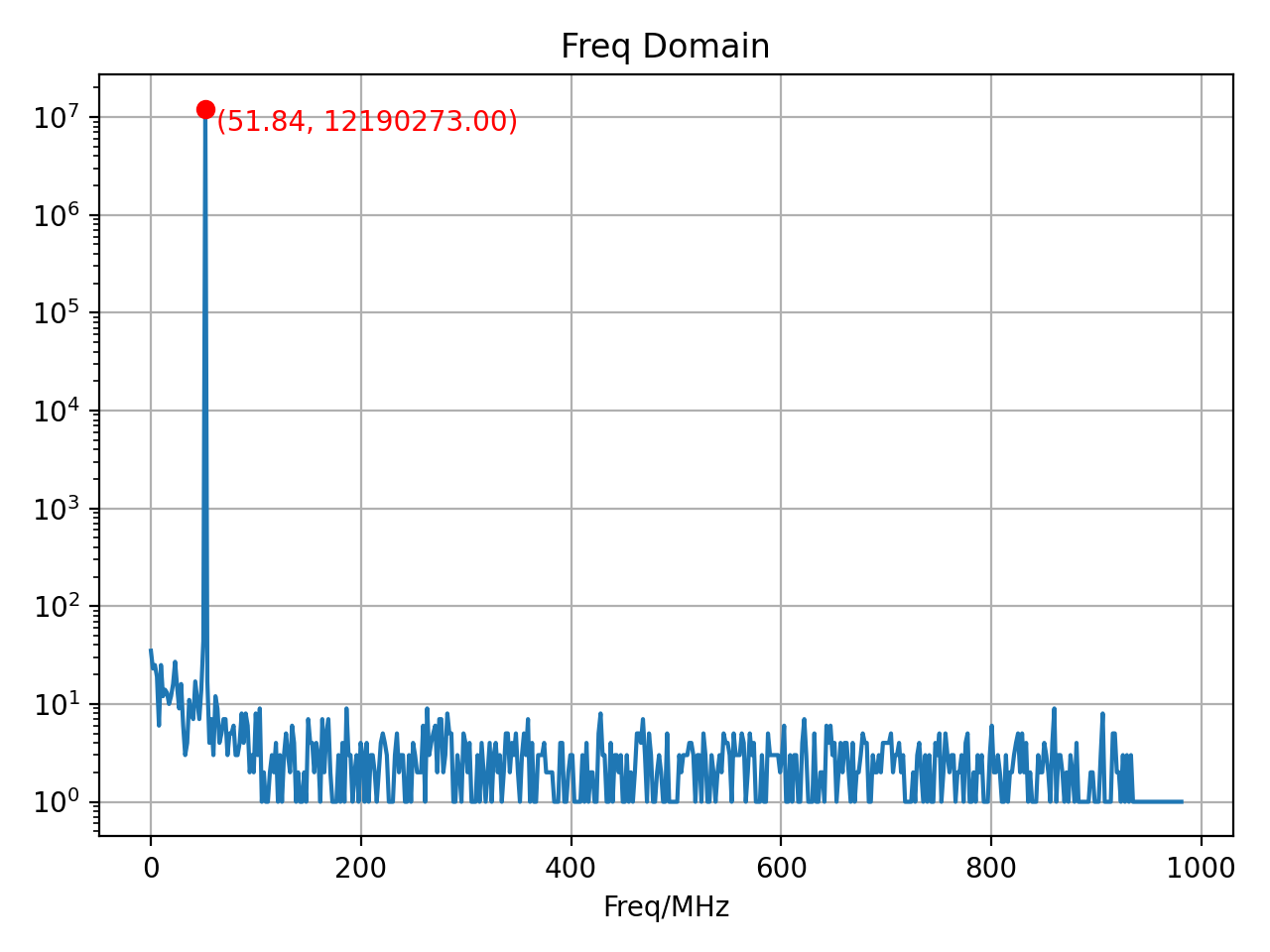}
        \label{scilab_spec_result}
    }
    \hspace{5pt}
    \subfloat[The spectra result from the demonstration design in Simulink.]{
        \includegraphics[height=6cm]{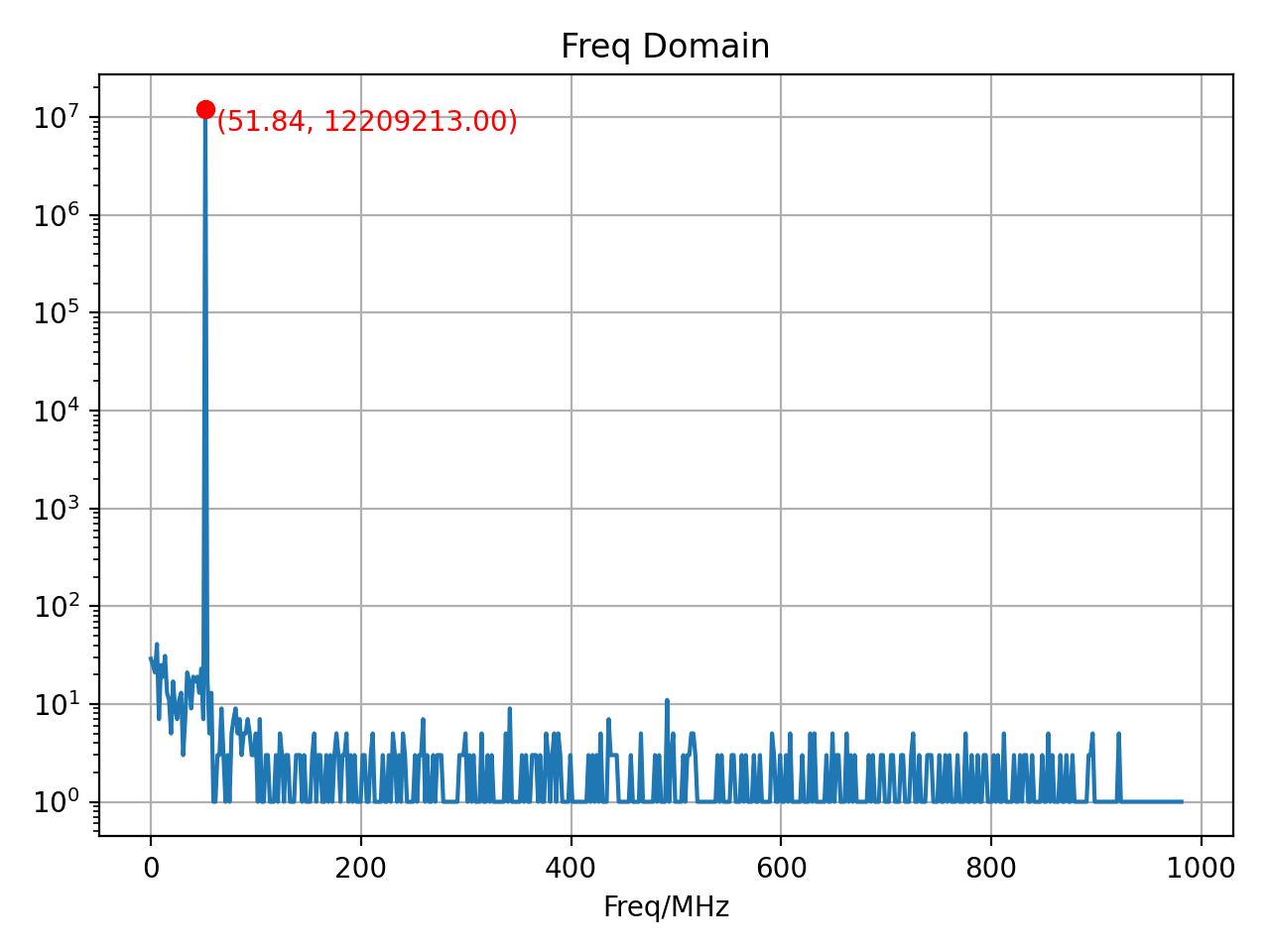}
        \label{simulink_spec_result}
    }\\
    \caption{The spectrometer design test results on RFSoC4x2.}
    \label{spectrometer_test_result}
\end{figure}

\indent The test results are depicted in Figure~\ref{spectrometer_test_result}. The spectrometer result from the Scilab demonstration design is illustrated in Figure~\ref{scilab_spec_result}, while the spectrometer result from the Simulink demonstration design is depicted in Figure~\ref{simulink_spec_result}. The two results are consistent to each other, which indicates that the spectrometer demonstration design functions on the RFSoC4x2 platform.\\

% \begin{figure}[htbp]
%     \centering
%     \includegraphics[height=6cm]{figures/spectrometer_test_result.png}
%     \caption{
%         The spectrometer design test result shows the signal is at 51.84MHz.
%         \label{spectrometer_test_result}
%     }
% \end{figure}

\section{Summary}
\label{summary}
\indent This paper introduces an open-source frontend, Scilab, for the CASPER toolflow. Certain Yellow blocks, such as RFDC, software register, and shared BRAM blocks, have been migrated to this new frontend along with several DSP blocks, including a Wideband FFT and an Accumulation block. A straightforward spectrometer demonstration is constructed using these blocks, and the test is conducted on an RFSoC4x2 board. The test results indicate that the CASPER toolflow utilizing Scilab offers nearly same functionalities to those found in MATLAB. The architecture for this new frontend is designed such that the porting additional alternative frontends to the CASPER toolflow should be feasible with minimal further changes. \\

\section{Acknowledgment}
The authors would like to thank Benjamin P Godfrey and Mitch Burnett for the discussion of the new toolflow development. This research was supported by NSF grant 2307781.

\bibliography{sample.bib}
\bibliographystyle{ws-jai}

\end{document}